\documentclass[sort&compress,final]{aipproc}\layoutstyle{6x9}
\begin{document}\title{Stability of Salpeter Solutions}
\classification{11.10.St, 03.65.Ge, 03.65.Pm}
\keywords{Bethe--Salpeter formalism, three-dimensional reduction,
instantaneous approximation, Salpeter equation, reduced Salpeter
equation, bound states in quantum field theory, stability
analysis}\author{Wolfgang LUCHA}{address={Institute for High
Energy Physics, Austrian Academy of Sciences,\\Nikolsdorfergasse
18, A-1050 Vienna, Austria}}\author{Franz
F.~SCH\"OBERL}{address={Faculty of Physics, University of Vienna,
Boltzmanngasse 5, A-1090 Vienna, Austria}}

\begin{abstract}By analytical spectral analysis the bound-state
solutions of the reduced Salpeter equation with
harmonic-oscillator interaction are shown to be free of the
instabilities numerically found in the (full) Salpeter equation
for confining Bethe--Salpeter interaction kernels of certain
Lorentz structure.\end{abstract}\maketitle

\section{Introduction: Motivation}In the framework of
`instantaneous approximations' to the Bethe--Salpeter formalism
for the description of bound states within quantum field theories,
depending on the Lorentz structure of the Bethe--Salpeter
interaction kernel solutions of the (full) Salpeter equation with
some confining interactions may exhibit certain instabilities
\cite{StabSal}, possibly related to the Klein paradox, signalling
the decay of states assumed to be bound by these confining
interactions, and observed in {\em numerical\/} (variational)
studies \cite{StabSal} of the Salpeter equation.

The (presumably) simplest scenario allowing for the {\em fully
analytic\/} investigation of this problem is set by the {\em
reduced\/} Salpeter equation \cite{Henriques76} with {\em
harmonic-oscillator\/} interaction. In this case, the integral
equation of Salpeter simplifies to either an algebraic relation or
a second-order homogeneous linear ordinary differential equation,
immediately accessible to standard techniques. There one can hope
to be able to decide unambiguously whether this setting poses a
well-defined (eigenvalue) problem the solutions of which
correspond to {\em stable\/} bound states associated to {\em
real\/} energy eigenvalues that are {\em bounded from below\/}.

\section{Reduced Salpeter Equation for Confining Interaction
Kernels of Harmonic-Oscillator Type}Assuming, as usual, the
Lorentz structures of the effective couplings of both fermion and
antifermion to be represented by identical matrices in Dirac
space, generically labeled~$\Gamma,$ and denoting the associated
Lorentz-scalar interaction function by
$V_\Gamma(\mathbf{p},\mathbf{q}),$ the so-called {\em reduced
Salpeter equation\/} \cite{Henriques76} describing bound states
composed of some fermion and its corresponding antifermion (with
common mass $m$ and relative momentum $\mathbf{p}$) reads for a
bound state with mass eigenvalue $M$ in the center-of-momentum
frame of the bound~state
\begin{equation}\displaystyle (M-2\,E)\,\Phi(\mathbf{p})=
\Lambda^+(\mathbf{p})\,\gamma_0\int\frac{{\rm
d}^3q}{(2\pi)^3}\,\sum_\Gamma V_\Gamma(\mathbf{p},\mathbf{q})\,
\Gamma\,\Phi(\mathbf{q})\,\Gamma\,
\Lambda^-(\mathbf{p})\,\gamma_0\ ,\label{Eq:RSE}\end{equation}with
the one-particle kinetic energy $E$ and energy projectors
$\Lambda^\pm(\mathbf{p})$ defined
according~to$$E\equiv\sqrt{p^2+m^2}\ ,\qquad
p\equiv|\mathbf{p}|\equiv\sqrt{\mathbf{p}^2}\
,\qquad\mbox{and}\qquad\Lambda^\pm(\mathbf{p})\equiv\frac{E\pm
\gamma_0\,(\mathbf{\gamma}\cdot\mathbf{p}+m)}{2\,E}\ .$$Any
solution to this eigenvalue problem is the Salpeter amplitude
$\Phi(\mathbf{p})$ of a bound state. Let the Bethe--Salpeter
kernel be of convolution type,
$V_\Gamma(\mathbf{p},\mathbf{q})=V_\Gamma(\mathbf{p}-\mathbf{q}),$
arising from a central potential $V(r),$ $r\equiv|\mathbf{x}|,$ in
configuration space. Then, for the harmonic-oscillator potential
$V(r)=a\,r^2,$ $a=a^\ast\ne0,$ the reduced~Salpeter equation
becomes a second-order differential equation utilizing the
Laplacian acting on states of angular momentum $\ell=0,$
$$D\equiv\frac{{\rm d}^2}{{\rm d}p^2}+\frac{2}{p}\,\frac{{\rm
d}}{{\rm d}p}\ .$$

In order to make contact with related previous analyses
\cite{Lucha00:IBSEm=0,Lucha00:IBSE-C4,Lucha00:IBSEnzm,Lucha01:IBSEIAS},
let us present our line of argument for fermion--antifermion bound
states of total spin $J,$ parity $P=(-1)^{J+1},$ and
charge-conjugation quantum number $C=(-1)^J,$ called ${}^1J_J$
spectroscopically. Due to the projectors $\Lambda^\pm(\mathbf{p})$
on the right-hand side of the reduced Salpeter equation
\eqref{Eq:RSE}, the Salpeter amplitudes $\Phi(\mathbf{p})$
describing these states contain only {\em one independent
component\/} $\phi(\mathbf{p})$:
$$\Phi(\mathbf{p})=2\,\phi(\mathbf{p})\,
\Lambda^+(\mathbf{p})\,\gamma_5\ .$$Somewhat more specifically, we
consider {\em pseudoscalar\/} (${}^1{\rm S}_0$) bound states,
characterized by the spin-parity-charge conjugation assignment
$J^{PC}=0^{-+}.$ A brief moment of thought reveals that the
cumbersome instabilities should appear first in the pseudoscalar
channel. Stripping off all dependence on angular variables
\cite{RadEVE} converts this {\em ``harmonic-oscillator reduced
Salpeter equation''\/} into the eigenvalue equation of a
(Hamiltonian) operator~${\cal H}$:\begin{equation}{\cal
H}\,\phi(p)=M\,\phi(p)\ .\label{Eq:EVE}\end{equation}

It is a straightforward task to work out the explicit form of all
the Hamiltonians $\cal H$ for the most popular choices of the
Lorentz structure of Bethe--Salpeter kernels (cf.\
Table~\ref{Tab:H}).

\begin{table}[h]\caption{Hamiltonian (differential or mere
multiplication) operators $\cal H$ entering into the eigenvalue
equation \eqref{Eq:EVE} equivalent to the~reduced Salpeter
equation \eqref{Eq:RSE} with harmonic-oscillator interaction
potential $V(r)=a\,r^2,$ for several frequently considered Lorentz
structures of the Bethe--Salpeter interaction
kernel.}\label{Tab:H}\begin{tabular}{ccc}\hline
\tablehead{1}{c}{b}{Lorentz structure}&
\tablehead{1}{c}{b}{$\Gamma\otimes\Gamma$}&
\tablehead{1}{c}{b}{$\cal H$}\\\hline\\[-2ex]Lorentz
scalar&$1\otimes1$&$\displaystyle
2\,E+a\left(\frac{2\,p^2+3\,m^2}{2\,E^4}
+\frac{m^2}{E}\,D\,\frac{1}{E}\right)$\\[2ex]time-component
Lorentz vector&$\gamma^0\otimes\gamma^0$&$\displaystyle
2\,E+a\left(\frac{2\,p^2+3\,m^2}{2\,E^4} -D\right)$\\[2ex] Lorentz
vector&$\gamma_\mu\otimes\gamma^\mu$&$\displaystyle
2\,E+a\left(\frac{m^2}{E}\,D\,\frac{1}{E}-2\,D\right)$\\[2ex]
Lorentz pseudoscalar&$\gamma_5\otimes\gamma_5$&$\displaystyle
2\,E+a\,\frac{2\,p^2+3\,m^2}{2\,E^4}$\\[1.5ex]B\"ohm--Joos--Krammer
(BJK) \cite{BJK73}&$\frac{1}{2}
\left(\gamma_\mu\otimes\gamma^\mu+\gamma_5\otimes\gamma_5-1\otimes1\right)$
&$\displaystyle 2\,E-a\,D$\\[1ex]\hline\end{tabular}\end{table}

\section{Spectral Properties of the Reduced-Salpeter Hamiltonian
Operators $\cal H$}The spectra of the multiplication or
differential operators $\cal H$ for various Dirac structures
$\Gamma\otimes\Gamma$ fixed by the Bethe--Salpeter kernel exhibit
the following stability-relevant features:\begin{itemize}\item All
of our {\em Hamiltonian operators\/} ${\cal H}$ are {\em
self-adjoint\/} since the differential operators $D$ and
$m^2\,E^{-1}\,D\,E^{-1}$ as well as the multiplication by any
real-valued function define self-adjoint operators. Consequently,
the entire {\em spectrum\/} of any operator $\cal H$ is {\em
real\/}. For ``reasonable'' interaction kernels, that is, for
kernels only constructed in terms of Dirac matrices $\Gamma$
subject to $\gamma_0\,\Gamma^\dag\,\gamma_0=\pm\Gamma$ and
potential functions $V_\Gamma(\mathbf{p},\mathbf{q})$ subject to
$V^\ast_\Gamma(\mathbf{q},\mathbf{p})=
V_\Gamma(\mathbf{p},\mathbf{q}),$ the {\em reality\/} of all {\em
eigenvalues\/} $M$ is guaranteed by the relation~\cite{RadEVE}
\begin{eqnarray*}&&M\int\frac{{\rm d}^3p}{(2\pi)^3}\,{\rm
Tr}\left[\Phi^\dag(\mathbf{p})\,\Phi(\mathbf{p})\right]
=2\int\frac{{\rm d}^3p}{(2\pi)^3}\,E\,{\rm
Tr}\left[\Phi^\dag(\mathbf{p})\,\Phi(\mathbf{p})\right]\\
&&+\int\frac{{\rm d}^3p}{(2\pi)^3}\int\frac{{\rm
d}^3q}{(2\pi)^3}\,\sum_\Gamma V_\Gamma(\mathbf{p},\mathbf{q})
\,{\rm Tr}\left[\Phi^\dag(\mathbf{p})\,\gamma_0\,\Gamma\,
\Phi(\mathbf{q})\,\Gamma\,\gamma_0\right],\end{eqnarray*}
satisfied by any Salpeter amplitude $\Phi(\mathbf{p})$ solving the
reduced Salpeter equation \eqref{Eq:RSE}.\item For the Lorentz
{\em pseudoscalar\/},
$\Gamma\otimes\Gamma=\gamma_5\otimes\gamma_5,$ and, if $m=0,$ for
the Lorentz {\em scalar\/}, $\Gamma\otimes\Gamma=1\otimes1,$ our
Hamiltonians ${\cal H}$ form pure multiplication operators, with
purely {\em continuous\/} spectrum. Bound states do not exist,
stability questions thus do not arise.\item For the {\em
time-component\/} Lorentz {\em vector\/},
$\Gamma\otimes\Gamma=\gamma^0\otimes\gamma^0,$ for the Lorentz
structure introduced in \cite{BJK73},
$\Gamma\otimes\Gamma=\frac{1}{2}\,
(\gamma_\mu\otimes\gamma^\mu+\gamma_5\otimes\gamma_5-1\otimes1),$
and, if $m=0,$ for the Lorentz {\em vector\/},
$\Gamma\otimes\Gamma=\gamma_\mu\otimes\gamma^\mu,$ our
Hamiltonians ${\cal H}$ form ($\ell=0$) Schr\"odinger operators
with a positive, infinitely rising potential $V(p)\to\infty$ for
$p\to\infty,$ provided, of course, the sign of the coupling $a$ is
chosen appropriately. The latter operators have entirely {\em
discrete\/} spectra bounded from below; all bound states may be
expected to be {\em stable\/}. Figure~\ref{Fig:V} illustrates the
typical qualitative behaviour of the effective potentials $V(p).$

\begin{figure}[h]\includegraphics[scale=1]{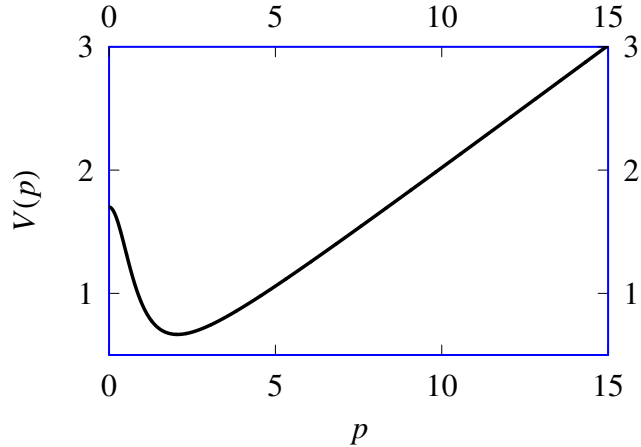}
\caption{Effective potential $V(p)$ in the Hamiltonian ${\cal H}$
for the harmonic-oscillator reduced Salpeter equation of
time-component Lorentz-vector kernel
$\Gamma\otimes\Gamma=\gamma^0\otimes\gamma^0,$ for $m=1$ and
$a=10$ [arbitrary~units].}\label{Fig:V}\end{figure}

\begin{figure}[hp]\begin{tabular}{c}
\includegraphics[height=.4\textheight,scale=1]{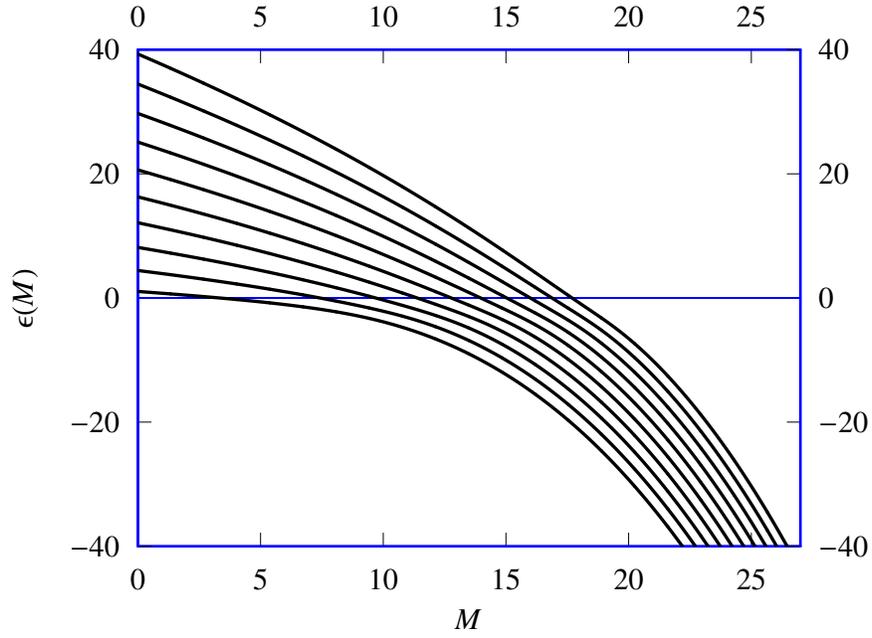}\\[0.05ex](a)\\[2ex]
\includegraphics[height=.4\textheight,scale=1]{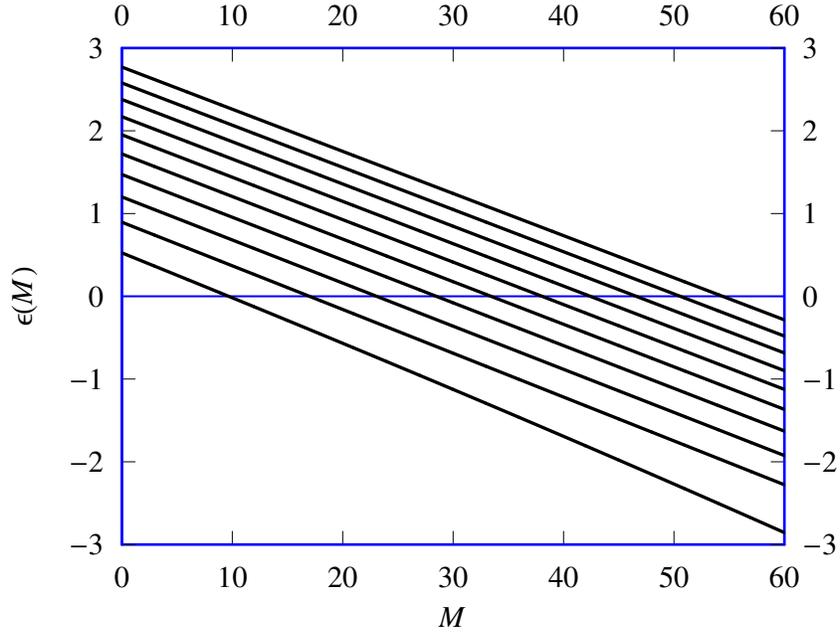}\\[0.05ex](b)\\[2ex]
\end{tabular}\caption{Lowest-lying eigenvalues $\epsilon(M)$ of the
auxiliary Hamiltonians ${\cal K}\equiv-D+U(p;M),$ derived from the
reduced Salpeter equation \eqref{Eq:RSE} describing pseudoscalar
(${}^1{\rm S}_0$) fermion--antifermion bound states of
constituents with mass $m=1,$ which experience a
harmonic-oscillator interaction [$V(r)=a\,r^2,$ $a=a^\ast\ne0$]
(a) of the \mbox{Lorentz-scalar} structure
$\Gamma\otimes\Gamma=1\otimes1$ (with the ``binding'' coupling
constant $a=-10$) and (b) of the Lorentz-vector structure
$\Gamma\otimes\Gamma=\gamma_\mu\otimes\gamma^\mu$ (with a
``binding'' coupling constant $a=10$) [arbitrary~units].}
\label{Fig:AEV}\end{figure}

\item For $m\ne0,$ because of the presence of the nasty
differential operators $m^2\,E^{-1}\,D\,E^{-1},$ the Hamiltonians
${\cal H}$ in \eqref{Eq:EVE} corresponding to both Lorentz {\em
scalar\/}, $\Gamma\otimes\Gamma=1\otimes1,$ and Lorentz {\em
vector\/}, $\Gamma\otimes\Gamma=\gamma_\mu\otimes\gamma^\mu,$ do
{\em not\/} constitute (standard) Schr\"odinger operators. In
these cases, however, by a suitable redefinition of the radial
amplitudes $\phi(p),$ the (radial) differential equations may be
transformed to eigenvalue equations of ($\ell=0$) Schr\"odinger
operators ${\cal K}$ of the form ${\cal K}\equiv-D+U(p;M),$ which
involve effective potentials $U(p;M)$ depending on the bound-state
mass $M$ as parameter. As might be guessed from the form of the
corresponding Hamiltonian ${\cal H},$ for the Lorentz {\em
scalar\/} the transformation required here simply reads
$\phi(p)\to E\,\phi(p).$ For given~$M,$ and the appropriate sign
of $a,$ our effective potentials $U(p;M)$ are bounded from below
and behave like $U(p;M)\to\infty$ for $p\to\infty.$ Thus, the
spectra of both ``auxiliary'' operators ${\cal K}$ must consist
entirely of {\em discrete\/} $M$-dependent eigenvalues
$\epsilon(M)$ (cf.\ Fig.~\ref{Fig:AEV}). The derivatives of the
latter eigenvalues with respect to $M$ are, for all $M,$ strictly
definite. The bound-state masses $M,$ defined by the zeros of the
eigenvalues $\epsilon(M),$ must then be discrete too. Since these
eigenvalues $\epsilon(M)$ are strictly {\em decreasing\/}
functions of $M$ a closer inspection establishes the bound-state
masses $M$ to be bounded from~{\em below\/}.\end{itemize}In
summary, given the semiboundedness of all the Hamiltonians ${\cal
H}$ entering in the radial equations and having established the
discreteness of their spectra for harmonic-oscillator couplings of
appropriate sign, our {\em harmonic-oscillator reduced Salpeter
equation\/} poses, at least for a very wide class of Lorentz
structures, a well-defined problem, with solutions describing {\em
stable bound states\/} related to some {\em real discrete spectra
bounded from below\/}.

\section{Generalization to the (Full) Salpeter Equation}It goes
without saying that a similar study may be envisaged for the {\em
full\/} Salpeter equation (as before, for bound-state constituents
of equal mass in the rest frame of the bound state)
\begin{eqnarray}\Phi(\mathbf{p})&=&\int\frac{{\rm d}^3q}{(2\pi)^3}
\left(\frac{\Lambda^+(\mathbf{p})\,\gamma_0\,\sum_\Gamma
V_\Gamma(\mathbf{p},\mathbf{q})\,\Gamma\,\Phi(\mathbf{q})\,\Gamma\,
\Lambda^-(\mathbf{p})\,\gamma_0}{M-2\,E}\right.\nonumber\\[1ex]
&&\hspace{7.88ex}\left.-\frac{\Lambda^-(\mathbf{p})\,\gamma_0\,\sum_\Gamma
V_\Gamma(\mathbf{p},\mathbf{q})\,\Gamma\,\Phi(\mathbf{q})\,\Gamma\,
\Lambda^+(\mathbf{p})\,\gamma_0}{M+2\,E}\right).\label{Eq:SE}\end{eqnarray}
There, however, the spectral analysis will be more complicated for
the following reasons:\begin{itemize}\item Although the {\em
squares\/} $M^2$ of all the mass eigenvalues $M$ are guaranteed to
be real \cite{Resag94} the spectrum is, in general, {\em not\/}
necessarily real and, even in those cases where it can be shown to
be real, it is certainly {\em not\/} bounded from below
\cite{Resag94}. In particular, for the perhaps most important
example of Bethe--Salpeter kernels involving only potential
functions $V_\Gamma(\mathbf{p},\mathbf{q})$ satisfying
$V^\ast_\Gamma(\mathbf{p},\mathbf{q})=V_\Gamma(\mathbf{p},\mathbf{q})
=V_\Gamma(\mathbf{q},\mathbf{p})$ and coupling matrices $\Gamma$
satisfying $\gamma_0\,\Gamma^\dag\,\gamma_0=\pm\Gamma$ the
spectrum of mass eigenvalues $M$ in the complex-$M$ plane consists
of {\em real\/} opposite-sign pairs $(M,-M)$ and {\em imaginary\/}
points $M=-M^\ast$.\item ``Full-Salpeter amplitudes''
$\Phi(\mathbf{p}),$ that is, solutions of the full Salpeter
equation \eqref{Eq:SE}, have more than one independent components.
Thus, any {\em full\/} Salpeter equation with harmonic-oscillator
interaction translates to a {\em system\/} of more than one
second-order differential equations or, equivalently, a single
differential equation of higher order.\end{itemize}The Salpeter
amplitude for ${}^1J_J$ states, e.g., involves two independent
components, $\phi_1,$~$\phi_2$:
$$\Phi(\mathbf{p})=\left[\phi_1(\mathbf{p})\,
\frac{\gamma_0\,(\mathbf{\gamma}\cdot\mathbf{p}+m)}{E}
+\phi_2(\mathbf{p})\right]\gamma_5\ .$$In particular, for
interaction kernels of Lorentz-scalar
($\Gamma\otimes\Gamma=1\otimes1$) or time-component Lorentz-vector
($\Gamma\otimes\Gamma=\gamma^0\otimes\gamma^0$) structure, the
{\em full\/} Salpeter equation \eqref{Eq:SE}, after getting rid of
all angular variables, becomes equivalent to the fourth-order
differential equation~\cite{HORSE}
\begin{eqnarray*}&&\left\{4\,E^2-\frac{2\,a}{E}
\left[\sigma\,(p^2+2\,m^2)\,D+p\,D\,p-2\right]
+\frac{a^2}{E}\left(m^2\,D+\sigma\,p\,D\,p-2\,\sigma\right)
\frac{1}{E}\,D\right\}\phi_2(p)\\&&\;=M^2\,\phi_2(p)\
,\end{eqnarray*}with a sign factor $\sigma$ (rendering possible a
simultaneous study of both structures) given~by
$$\sigma=\left\{\begin{array}{ll}+1\qquad\mbox{for
$\Gamma\otimes\Gamma=\gamma^0\otimes\gamma^0$}&\mbox{(time-component
Lorentz-vector interactions)}\ ,\\[1ex]-1\qquad\mbox{for
$\Gamma\otimes\Gamma=1\otimes1$}&\mbox{(Lorentz-scalar
interactions)}\ .\end{array}\right.$$A notable exception is the
Salpeter equation \eqref{Eq:SE} with a harmonic-oscillator
interaction~of the BJK \cite{BJK73} Lorentz structure
$\Gamma\otimes\Gamma=\frac{1}{2}
\left(\gamma_\mu\otimes\gamma^\mu+\gamma_5\otimes\gamma_5-1\otimes1\right)$:
here one still arrives at a second-order differential equation
that can be expressed in two equivalent ways \cite{HORSE},
$$4\left(E^2-a\,D\,E\right)\phi_1(p)=M^2\,\phi_1(p)\qquad\mbox{or}\qquad
4\left(E^2-a\,E\,D\right)\phi_2(p)=M^2\,\phi_2(p)\ ,$$since
$2\,E\,\phi_1(p)=M\,\phi_2(p).$ Again, the spectrum of bound-state
masses $M$ is discrete~\cite{HORSE}.

\begin{theacknowledgments}We wish to express our deepest gratitude
to Bernhard Baumgartner, Harald Grosse and Heide Narnhofer for
many interesting, stimulating, encouraging and helpful
discussions.\end{theacknowledgments}

\bibliographystyle{aipproc}\end{document}